\newcommand{\beq}{\begin{equation}}
\newcommand{\eeq}{\end{equation}}
\newcommand{\bqa}{\begin{eqnarray}}
\newcommand{\eqa}{\end{eqnarray}}
\newcommand{\beqa}{\begin{eqnarray}}
\newcommand{\eeqa}{\end{eqnarray}}
\newcommand{\beqan}{\begin{eqnarray*}}
\newcommand{\eeqan}{\end{eqnarray*}}
\newcommand{\nn}{\nonumber}
\newcommand{\erf}[1]{(\ref{#1})}

\newcommand{\bra}[1]{\langle{#1}|}
\newcommand{\ket}[1]{|{#1}\rangle}
\newcommand{\abra}[1]{\langle\underline{#1}|}
\newcommand{\aket}[1]{|\underline{#1}\rangle}
\newcommand{\ip}[1]{\langle{#1}\rangle}
\newcommand{\A}{{\rm A}}
\newcommand{\B}{{\rm B}}
\newcommand{\id}{{{\bf 1}}}

\newcommand{\refcite}[1]{{\cite{#1}}}

\documentclass[twocolumn,showpacs,preprintnumbers,pra,showkeys,nofootinbib]{revtex4}

\begin{document}



\title{Entanglement of identical particles and reference phase
uncertainty}

\author{
John A.Vaccaro}
\address{
    Quantum Physics Group, STRC,
    University of Hertfordshire, College Lane, Hatfield AL10 9AB, U.K.}
\author{
Fabio Anselmi}
\address{
    Quantum Physics Group, STRC,
    University of Hertfordshire, College Lane, Hatfield AL10 9AB, U.K.}
\author{
Howard M. Wiseman}
\address{
   Centre for Quantum Computer Technology, Centre for Quantum
   Dynamics, School of Science, Griffith University, Brisbane,
   Queensland 4111 Australia. }



\begin{abstract}
We have recently introduced a measure of the bipartite
entanglement of identical particles, $E_{\rm P}$, based on the
principle that entanglement should be {\em accessible} for use as
a resource in quantum information processing. We show here that
particle entanglement is limited by the lack of a reference phase
shared by the two parties, and that the entanglement is
constrained to reference-phase invariant subspaces.  The
super-additivity of $E_{\rm P}$ results from the fact that this
constraint is weaker for combined systems. A shared reference
phase can only be established by transferring particles between
the parties, that is, with additional nonlocal resources. We show
how this nonlocal operation can increase the particle
entanglement.
\end{abstract}


\maketitle

\section{Introduction}

Entanglement is an essential resource for quantum information
processing. The non separability of the wavefunction of two
distinct systems is the usual hallmark of an entangled state.
However, the symmetric or antisymmetric wavefunctions of
collections of identical particles is {\em inherently} non
separable. A crucial question then is how to quantify the
entanglement of identical particles. The approach of Zanardi and
others\cite{mode_ent} is to calculate the entanglement of the
quantum field modes, $E_{\rm M}$, rather than the particles that
occupy them. In particular, $E_{\rm M}$ can be non-zero even for
the case of a single particle. An alternate
approach\cite{quan_corr} is to examine the non separability of the
wavefunction beyond that required by symmetrization or
antisymmetrization.  The difficulty here, however, is that there
is no fixed partition into distinct systems.

The approach we take\cite{WiseVacc} is to insist that the
entanglement of the particles, $E_{\rm P}$, is {\em accessible} in
the sense that it could be transferred to regular quantum
registers (e.g. qubit systems) using local operations;  once
transferred it can be used as a generic resource for quantum
information processing. This requires strict {\em partite
separation} and the entanglement to be accessible using {\em local
operations} only. Transporting particles between the parties is
clearly a nonlocal operation; local operations therefore preserve
the local particle number at each site. Hence, these restrictions
are equivalent to imposing a local particle number superselection
rule.\cite{WiseVacc} Entanglement constrained by general
superselection rules have been explored further by Bartlett and
Wisemen.\cite{BartWise} A more introductory treatment can be found
in Ref.~\refcite{WisBarVac03}.  Also the impact of superselection
rules on nonlocality and quantum data hiding has been studied by
Verstraete and Cirac.\cite{VerCir}

While the local transfer of the particle entanglement to regular
quantum registers underpins our definition of $E_{\rm P}$ in Ref.\
\refcite{WiseVacc}, we did not explicitly show how the transfer
operation might be implemented. In this paper\cite{Kyoto} we give
an explicit demonstration of the transfer. We show how the lack of
a shared reference phase reduces the entanglement of the regular
quantum registers to that of $E_{\rm P}$. Moreover, by performing
a measurement of the difference between the reference phases at
the two sites, it is possible to recover the entropy of
entanglement of the original system. However, this requires the
transport of particles from one site to the other. The essential
point is that the entanglement can be recovered only by violating
local particle conservation and transporting particles from one
site to the other, that is, only by the use of other nonlocal
resources. We establish a relation between the variance in the
number of particles transported and the amount of entanglement in
the quantum registers. This explicit demonstration gives further
insight into the nature of particle entanglement and reference
phase uncertainty.

Reference phase uncertainty has recently been discussed in
relation to continuous variable teleportation\cite{cv_tele} and
communication without a shared reference frame.\cite{BartRudo} The
close connection between the application of local superselection
rules and a reference system has been discussed recently by Kitaev
{\em et al.}\cite{Kitaev} They show how one can simulate the local
violation of a superselection rule if a shared reference system is
available. Their analysis is in the context of data security
whereas our work here explores the implications for particle
entanglement. Schuch {\em et al.}\cite{Schuch} have recently
investigated the intra-conversion of sets of states under the
constraint of the local superselection rule associated with
particle conservation. They identify a new nonlocal resource
corresponding to the variance in local particle number. The
connection with our work is that we give explicit protocols for
converting this resource into particle entanglement by
establishing a shared reference frame.

The body of the paper is organized as follows.  We begin in
Section \ref{review} with a brief review the definition of $E_{\rm
P}$. In Section \ref{protocol} we describe the protocol for
transferring the entanglement of shared particles to regular
quantum registers for a variety of cases.  In Section
\ref{ref_phase} we show how the particle entanglement can be
increased by establishing a shared reference phase.  We end with a
discussion in Section \ref{discussion}.

\section{Entanglement of identical particles} \label{review}

We imagine two well-separated parties, Alice and Bob, sharing a
collection of $N$ identical particles, such as atoms or electrons
etc., which are in the pure state $\ket{\Psi}_{\rm AB}$. The {\em
particle entanglement} $E_{\rm P}(\ket{\Psi}_{\rm AB})$ of this
state is the maximum entanglement that can be transferred to local
quantum registers {\em without additional nonlocal resources}. We
showed in Ref.\ \refcite{WiseVacc} that this is given by:
\beq
   E_{\rm P}(\ket{\Psi}_{\rm AB}) \equiv
     \sum_{n=0}^N P_n E(\ket{\Psi_n}_{\rm AB})
     \label{E_P}
\eeq
where
\beq
   \ket{\Psi_n}_{\rm AB} = \frac{\hat{\Upsilon}_{n}
             \ket{\Psi}_{\rm AB}}{\sqrt{P_n}}\ ,
\eeq
$\hat{\Upsilon}_n$ is the projector onto states with $n$ particles
at Alice's site and $N-n$ at Bob's, $P_n = {}_{\rm
AB}\ip{\Psi|\hat{\Upsilon}_{n}|\Psi}_{\rm AB}$ is the probability
of finding $n$ particles at Alice's site, $\ket{\Psi_n}_{\rm AB}$
represents field modes occupied by a fixed number of particles at
each site, $E(\ket{\Psi_n}_{\rm AB})= S(\hat\rho_{\rm A}^{(n)})$
is the entropy of entanglement in $\ket{\Psi_n}_{\rm AB}$,
$S(\hat\rho)$ is the binary von Neumann entropy $-{\rm
Tr}(\hat\rho \log_{2}\hat\rho)$, and $\hat\rho_{\rm A}^{(n)}$ is
the reduced density matrix $\hat\rho_{\rm A}^{(n)} = {\rm
Tr}_{B}[(\ket{\Psi_n}\bra{\Psi_n})_{\rm AB}]$.\footnote{To
simplify the notation for projectors, we write the subscript AB
outside a bracket, e.g. as $(\ket{\psi}\bra{\psi})_{\rm AB}$,
rather than individually on each bra and ket.} In essence,
\erf{E_P} results from a {\em local} particle number
superselection rule in that the coherences between subspaces of
differing local particle number are not observable by local means.

In the following section we demonstrate the transfer of the
entanglement in $\ket{\Psi}_{\rm AB}$ to regular quantum
registers.  The essential features this operation are clearly
revealed in the simplest system: coherently sharing a single
particle between Alice and Bob in the state
\beq
  \ket{\Psi^{(1)}}_{\rm AB}=\frac{1}{\sqrt{2}}(\ket{1,0}_{\rm AB}+\ket{0,1}_{\rm
  AB})
  \label{shared_particle}
\eeq
where $\ket{i,j}_{\rm AB}$ represents $i$ particles in a field
mode at Alice's site and $j$ particles in a field mode at Bob's
site. We note, in particular, that sharing a single particle and
independently sharing two particles carries the following particle
entanglement:\cite{WiseVacc}
\beqa
  E_{\rm P}(\ket{\Psi^{(1)}}_{\rm AB})&=&0 \label{Ep_1}\\
  E_{\rm P}(\ket{\Psi^{(1)}}_{\rm AB}\otimes\ket{\Psi^{(1)}}_{\rm AB})
      &=&\frac{1}{2}\ .
  \label{Ep_2}
\eeqa
This illustrates a striking general feature of $E_{\rm P}$ in that
it is super-additive.  The super-additivity is a direct
consequence of the inherent indistinguishability of the particles.

\section{Transfer protocol and reference phase uncertainty}
\label{protocol}

We now demonstrate the transfer protocol of the particle
entanglement to regular qubit registers. We treat explicitly the
case of bosons here; the modification required for the fermion
case is, however, straightforward.\footnote{For the fermion case
we replace the state of a $n$-boson occupied mode $\ket{n}$ with
the state representing $n$ fermions distributed in $n$ different
modes each of which contain a single fermion: $ \ket{n}^{(f)}
\equiv \ket{1,1,\cdots,1,0,\cdots,0}$ where the number of modes
exceeds the number of fermions.  The protocol then involves
operations of the same form as the boson case.} Let Alice have a
very large number $M\gg 1$ of identical ancillary particles in a
particular field mode, i.e. the mode occupation is given by the
state $\ket{M}_\A$. An operation is then performed which shares
the particles with another mode at Alice's site to produce the
state
\beq
    \sum_{n=0}^M c_n\ket{M-n,n}_{\rm A}
\eeq
where here $\ket{i,j}_\A$ represents $i$ particles in one field
mode and $j$ particles in the second field mode at Alice's site,
and $c_n$ are complex amplitudes satisfying $\sum_n |c_n|^2=1$. We
can rewrite this state as
\beq
  \frac{\sqrt{M+1}}{2\pi}\int_{2\pi}\ket{c(\theta)}_{\rm A}
           \ket{\psi(\theta)}_{\rm A}d\theta
  \label{super_1}
\eeq
where
\beqa
   \ket{\psi(\theta)}  &=& \sum_{n=0}^M
                             \frac{e^{-in\theta}}{\sqrt{M+1}}\ket{M-n}
                          = \sum_{n=0}^M
                             \frac{e^{-i(M-n)\theta}}{\sqrt{M+1}}\ket{n}\
                             \
                             \label{trunc_phase}\\
    \ket{c(\theta)} &=& \sum_{n=0}^M
                            c_n e^{in\theta}\ket{n}\ .
                            \label{c_state}
\eeqa
Here $\ket{\psi(\theta)}$ is a ``truncated'' phase
state\cite{PB,London} and $\ket{c(0)}$ is a state with a large
mean particle number $\overline{N}_c=\sum_n |c_n|^2 n$ satisfying
$M\gg \overline{N}_c\gg 1$, but otherwise arbitrary. For example,
$\ket{c(\theta)}$ could approximate a large amplitude coherent
state with $c_n\propto({\overline{N}_c^n
e^{-\overline{N}_c}}/{n!})^{1/2}$ .  A corresponding process is
performed at Bob's site with his local ancillary system, resulting
in the combined ancillary state
\beq
  \frac{M+1}{(2\pi)^2}\int_{2\pi}\ket{c(\theta)}_{\rm A}
      \ket{\psi(\theta)}_{\rm A}d\theta
  \int_{2\pi}\ket{c(\phi)}_{\rm B}
       \ket{\psi(\phi)}_{\rm B}d\phi
  \ .
  \label{super_m}
\eeq

\subsection{Single shared particle}

Our transfer protocol is based on a method introduced by
Mayers.\cite{mayers}  We demonstrate it first for the state of a
single shared particle, $\ket{\Psi^{(1)}}_{\rm AB}$ in
\erf{shared_particle}.  Let the initial state of the two regular
qubits, one at Alice's site and the other at Bob's, be
$\aket{0,0}_{\rm AB}$. We use an underline to distinguish the
states of a regular qubit, $\aket{0}$, $\aket{1}$, (such as two
orthogonal electronic states of an atom) from the Fock states of
the field modes $\ket{0}$, $\ket{1}$, $\cdots$.

We will first concentrate on the integrand of the left integral in
\erf{super_m} for a specific value of $\theta$. Let this term
together with the shared particle modes and a single regular qubit
at Alice's site be given by
\beqa
   &&\cdots\ket{\psi(\theta)}_\A\otimes
   \ket{\Psi^{(1)}}_{\rm AB}\otimes\aket{0}_\A\nonumber\\
   &&=\cdots\ket{\psi(\theta)}_\A\otimes
   \frac{1}{\sqrt{2}}\left(\ket{1,0}_{\rm AB}
   +\ket{0,1}_{\rm AB}\right)\otimes\aket{0}_A\ \
\eeqa
where, for clarity, we have reordered the states and written
``$\cdots$'' to represent states of modes that are not of
immediate interest. Alice performs a local CNOT operation using
her local shared-particle mode as the control and her local
regular qubit as the target, yielding
\beq
   \cdots\ket{\psi(\theta)}_\A\otimes
   \frac{1}{\sqrt{2}}\left(\ket{1,0}_{\rm AB}\otimes\aket{1}_\A
   +\ket{0,1}_{\rm AB}\otimes\aket{0}_\A\right)\ .
\eeq
To complete her part of the protocol, Alice must disentangle her
shared-particle mode from her regular qubit for this value of
$\theta$. This entails ``hiding'' the shared particle in the
truncated phase state $\ket{\psi(\theta)}_\A$. Expanding the state
$\ket{\psi(\theta)}_\A$ in the number basis yields
\beqa
   \cdots\frac{1}{\sqrt{2(M+1)}}\sum_{n=0}^M
      &e^{-i(M-n)\theta}\Big[\ket{n}_\A
          \otimes\ket{1,0}_{\rm AB}\otimes\aket{1}_\A\nn\\
      &+\ket{n}_\A\otimes\ket{0,1}_{\rm AB}
               \otimes\aket{0}_\A\Big]\ .
\eeqa
Alice now applies a controlled operation with her regular qubit as
the control and the mapping:
\begin{widetext}
\beq
   \cdots\ket{x}_\A\otimes\ket{y,z}_{\rm
   AB}\otimes\aket{0}_\A
      \mapsto  \cdots\ket{x}_\A\otimes\ket{y,z}_{\rm
          AB}\otimes\aket{0}_\A\ ,
          \label{control1}
\eeq
\beq
   \cdots\ket{x}_\A\otimes\ket{y,z}_{\rm
   AB}\otimes\aket{1}_\A
       \mapsto  \cdots\ket{x+y}_\A\otimes
          \ket{0,z}_{\rm AB}\otimes\aket{1}_\A\ ,
          \label{control2}
\eeq
to produce the state
\beq
   \cdots\frac{1}{\sqrt{2}}\Big\{\ket{\psi(\theta)}_\A\otimes
   \left(e^{-i\theta}\ket{0,0}_{\rm AB}\otimes\aket{1}_\A
   +\ket{0,1}_{\rm AB}\otimes\aket{0}_\A\right)
    +\frac{1}{\sqrt{M+1}}\left[\ket{M+1}_\A-e^{-i(M+1)
   \theta}\ket{0}_\A\right]\otimes\ket{0,0}_{\rm AB}\otimes\aket{1}_\A\Big\}\
   .
\eeq

Next Bob repeats these operations using his truncated phase state
and another regular qubit in the state $\aket{0}_\B$ at his site
as follows. We first consider the integrand of the right integral
in \erf{super_m} for a specific value of $\phi$.  We also reorder
the states and include only states of modes that are of immediate
interest:
\beqa
   &&\cdots\frac{1}{\sqrt{2}}\Big\{\ket{\psi(\theta)}_\A\otimes
   \left(e^{-i\theta}\ket{0,0}_{\rm AB}\otimes\aket{1}_\A
   +\ket{0,1}_{\rm AB}\otimes\aket{0}_\A\right)\nonumber\\
   &&+\frac{1}{\sqrt{M+1}}\left[\ket{M+1}_\A-e^{-i(M+1)
   \theta}\ket{0}_\A\right]\otimes\ket{0,0}_{\rm AB}\otimes\aket{1}_\A\Big\}
   \otimes\aket{0}_\B\otimes\ket{\psi(\phi)}_\B\ .\ \ \ \ \
\eeqa
Bob performs a local CNOT operation using his local
shared-particle mode as the control and his regular qubit as the
target. He then performs a controlled operation analogous to
\erf{control1} and \erf{control2} using his regular qubit as the
control and his truncated phase state $\ket{\psi(\phi)}_\B$ as the
target. This gives the state
\beqa
   &&\cdots\frac{1}{\sqrt{2}}\Big\{\ket{\psi(\theta)}_\A\otimes
   \ket{0,0}_{\rm AB}\otimes\left(e^{-i\theta}\aket{1,0}_{\rm AB}
   +e^{-i\phi}\aket{0,1}_{\rm AB}\right)\otimes\ket{\psi(\phi)}_\B
   \nonumber\\
   &&+\frac{1}{\sqrt{M+1}}\ket{\psi(\theta)}_\A\otimes\ket{0,0}_{\rm AB}
   \otimes\aket{0,1}_{\rm AB}\otimes
   \left[\ket{M+1}_\B-e^{-i(M+1)\phi}\ket{0}_\B\right]\nonumber\\
   &&+\frac{1}{\sqrt{M+1}}\left[\ket{M+1}_\A-e^{-i(M+1)\theta}
   \ket{0}_\A\right]\otimes\ket{0,0}_{\rm AB}\otimes\aket{1,0}_{\rm AB}
   \otimes\ket{\psi(\phi)}_\B\Big\}\ .\label{trans1}
\eeqa
Here, and in the following, we write the state
$\aket{n}_\A\otimes\aket{m}_\B$ of the regular qubits as
$\aket{n,m}_{\rm AB}$, for convenience. For the limiting case of
large $M$ the statistical weighting of the last two terms, being
of order $2/(M+1)$, becomes vanishingly small.  We ignore these
terms for the remainder of this paper.  We now trace over all
particle modes as our interest lies only in the regular qubits.
Recalling that the state being considered is part of the
integrands in \erf{super_m}, we find that we need to evaluate
integrals of the following form
\beq
   I=\int_{2\pi}\sum_{n=0}^M\sum_{m=0}^M\ip{n|\psi(\theta)}\ip{\psi(\theta')|n}
      \ip{m|c(\theta)}\ip{c(\theta')|m}e^{ik\theta'}\frac{d\theta'}{2\pi}
      \label{I1}
\eeq
\end{widetext}
where $k$ is a non-negative integer.  Using the expansions of
$\ket{\psi(\theta)}$  and $\ket{c(\theta)}$ in terms of the Fock
states in \erf{trunc_phase} and \erf{c_state} shows that this
expression is simply
\beq
   I=\sum_{m=k}^M\frac{|c_m|^2}
        {(M+1)}e^{ik\theta}\approx\frac{1}{M+1}e^{ik\theta}\ .
        \label{I2}
\eeq
We have assumed here that the state $\ket{c(\theta)}$ has
negligible overlap with $\ket{n}$ for $n\le k$; this is the case,
for example, if $\ket{c(\theta)}$ approximates a large amplitude
coherent state with $|c_n|^2\propto{\overline{N}_c^n
e^{-\overline{N}_c}}/{n!}$ . Armed with this result, we find that
the qubit registers on their own are left in the mixed state:
\beqa
   &&\frac{1}{2}\int_{2\pi}\int_{2\pi}
      \left(e^{-i\theta}\aket{1,0}_{\rm AB}+
         e^{-i\phi}\aket{0,1}_{\rm AB}\right)\nn\\
    &&\ \ \ \ \ \quad\times  \left({}_{\rm AB}\abra{1,0}e^{i\theta}+
        {}_{\rm AB}\abra{0,1}e^{i\phi}\right)
           \frac{d\theta}{2\pi}\frac{d\phi}{2\pi}\nonumber\\
   &&=\frac{1}{2}\left(\aket{1,0}\abra{1,0}
         +\aket{0,1}\abra{0,1}\right)_{\rm AB}
        \ .   \label{mixed1}
\eeqa
As predicted in Ref.\ \refcite{WiseVacc} and shown in \erf{Ep_1},
there is no entanglement here. {\em The origin of the loss of
entanglement can therefore be attributed to the unknown phase
difference $\theta-\phi$ that emerges in the transfer protocol,
i.e. to the lack of a shared reference phase.}

\subsection{Independently sharing two particles}

This situation can be contrasted with the result of independently
sharing two particles, that is when Alice and Bob share the state
\beqa
   &&\ket{\Psi^{(1)}}_{\rm AB}\otimes\ket{\Psi^{(1)}}_{\rm AB}\nn\\
   &&=\frac{1}{\sqrt{2}}(\ket{1,0}_{\rm AB}+\ket{0,1}_{\rm
  AB})\otimes\frac{1}{\sqrt{2}}(\ket{1,0}_{\rm AB}+\ket{0,1}_{\rm
  AB})\nn\\
\eeqa
Carrying out the above transfer operations on the first shared
particle results in the state represented by the first line of
\erf{trans1} with probability $P=1-2/(M+1)$:
\beqa
   &&\cdots\frac{1}{\sqrt{2}}\ket{\psi(\theta)}_\A\otimes
   \ket{0,0}_{\rm AB}\nn\\
   &&\ \otimes\left(e^{-i\theta}\aket{1,0}_{\rm AB}
   +e^{-i\phi}\aket{0,1}_{\rm AB}\right)\otimes\ket{\psi(\phi)}_\B
   \ .
\eeqa
Repeating the operations on the second shared particle using the
truncated phase states $\ket{\psi(\theta)}_\A$ and
$\ket{\psi(\phi)}_\B$ and two additional regular qubits (one at
Alice's site and the other at Bob's) results in the reduced
density operator for the four regular qubits as
\begin{widetext}
\beqa
   &&\int_{2\pi}\int_{2\pi}
        \Big[\ket{R(\theta,\phi)}\otimes\ket{R(\theta,\phi)}\Big]\
   \Big[\bra{R(\theta,\phi)}\otimes\bra{R(\theta,\phi)}\Big]
    \frac{d\theta}{2\pi}\frac{d\phi}{2\pi}\nonumber\\
   &&=\frac{1}{4}\Big[\aket{00,11}\abra{00,11}
       +  \aket{11,00}\abra{11,00}
    +  \left(\aket{10,01}
             +\aket{01,10}\right)
           \left(\abra{10,01}+\abra{01,10}\right)\Big]_{\rm AB}
   \label{mixed2}
\eeqa
\end{widetext}
where
\beqa
   \ket{R(\theta,\phi)}=\frac{1}{\sqrt{2}}\left(e^{-i\theta}\aket{1,0}_{\rm AB}+
   e^{-i\phi}\aket{0,1}_{\rm AB}\right)
\eeqa
and we have written the joint state $\aket{i,j}_{\rm
AB}\otimes\aket{n,m}_{\rm AB}$ as $\aket{in,jm}_{\rm AB}$. Each of
the parties, Alice and Bob, can perform a local measurement to
determine if the states of their two regular qubits at their site
are equal or different; the result of the measurement is equally
likely to be
\beq
    \frac{1}{2}\Big(\aket{00,11}\abra{00,11} +  \aket{11,00}\abra{11,00}\Big)_{\rm AB}
\eeq
or
\beq
   \frac{1}{2}\Big[\left(\aket{10,01}+\aket{01,10}\right)
           \left(\abra{10,01}+\abra{01,10}\right)\Big]_{\rm AB}\ ,
     \label{ent}
\eeq
respectively.  The entanglement in the first result is zero
whereas it is 1 ebit in the second, and so the average
entanglement is 1/2 ebit. This agrees exactly with \erf{Ep_2} and
Ref.\ \refcite{WiseVacc}.

We note that the subspace spanned by the states
$\{\aket{10,01}_{\rm AB},\aket{01,10}_{\rm AB}\}$ is invariant to
arbitrary shifts of the local reference phases.  For example, the
reference phase shifts given by $\aket{0}_\A \mapsto \aket{0}_\A$,
$ \aket{1}_\A \mapsto e^{i\theta}\aket{1}_\A$, $\aket{0}_\B
\mapsto \aket{0}_\B$, $\aket{1}_\B \mapsto e^{i\phi}\aket{1}_\B$
map an arbitrary state of this subspace, $\alpha\aket{10,01}_{\rm
AB}+\beta\aket{01,10}_{\rm AB}$, to the state
$e^{i(\theta+\phi)}(\alpha\aket{10,01}_{\rm
AB}+\beta\aket{01,10}_{\rm AB})$, which differs only by an overall
phase factor from the original state. Clearly, in the absence of a
shared reference phase, the transferred entanglement is
constrained to such {\em reference-phase invariant subspaces}.
Comparing \erf{mixed1} and \erf{ent} we conclude that {\em the
super-additivity of $E_{\rm P}$ is due to this constraint being
weaker for the combined system}.

\subsection{The general case}

The transfer protocol can easily be generalized to multi-occupied
field modes where the $n$-particle state $\ket{n}$ is mapped to
the state $\aket{n}$ of a regular quantum register. Here
$\{\aket{n}:n=0,1,\cdots\}$ is an orthogonal basis set. We write a
general pure state representing $N$ particles shared between Alice
and Bob as
\beq
  \ket{\Psi}_{\rm AB}=\sum_{n=0}^N g_n\ket{\psi_n}_{\rm AB}
  \label{generalN}
\eeq
where $g_n$ are complex amplitudes and $\ket{\psi_n}_{\rm AB}$
represents a state comprising $n$ particles at Alice's site and
the remainder at Bob's site.  In \ref{gen_protocol} we show that
the final state of the regular quantum registers after the
transfer protocol is
\beq
   \sum_{n=0}^N |g_n|^2\Big(\aket{\psi_n}\abra{\psi_n}\Big)_{\rm AB}
   \label{mixedN}
\eeq
where $\aket{\psi_n}_{\rm AB}$ is the regular quantum register
version of the shared particles state $\ket{\psi_n}_{\rm AB}$.
Each of the terms in the sum of \erf{mixedN} belongs to a
different reference-phase invariant subspace. It is possible to
make a local measurement which projects onto these subspaces. Thus
the entanglement of \erf{mixedN} is given by \erf{E_P} with
$\ket{\psi_n}$ replaced by $\aket{\psi_n}$.  In other words, the
entanglement transferred to the quantum registers is in exact
agreement with our definition of $E_{\rm P}$. Moreover, this shows
that the transferred entanglement is constrained to
reference-phase invariant subspaces, in general.

\section{Entanglement and reduced reference phase uncertainty}
\label{ref_phase}

The foregoing suggests that the particle entanglement can be
increased by fixing the phase difference between the two sites.
Indeed, consider the case of sharing a single particle which
results in the mixed state in \erf{mixed1} in the absence of a
known phase difference $\theta-\phi$. The entanglement in this
state can be increased if we reduce the uncertainty in the phase
difference so that the implicit phase distributions in the
integral in \erf{mixed1} are no longer flat.  One way of doing
this is to perform a phase-difference measurement between the two
sites.  In \ref{phase_meas} we show that the state of the regular
qubits following an ideal phase-difference measurement of the
ancillary states $\ket{c(\theta)}_\A\otimes\ket{c(\phi)}_\B$ in
\erf{super_m} is given by
\beq
  \frac{1}{2}\Big(\aket{1,0}\abra{1,0}
  +C\aket{1,0}\abra{0,1}
             +C^*\aket{0,1}\abra{1,0}
   +\aket{0,1}\abra{0,1}\Big)_{\rm AB}
   \label{mixedC}
\eeq
where
\beq
  C=\int_{2\pi}\int_{2\pi}P_\varphi(\theta,\phi)e^{i(\phi-\theta)}d\theta
  d\phi
  \label{C}
\eeq
and $P_\varphi(\theta,\phi)$, which is defined in \erf{P_varphi},
represents the resolution of the phase-difference measurement for
a measured difference of $\varphi$ . The entanglement of
formation\cite{EF} of the mixed state \erf{mixedC} is given by
\beq
  E_{\rm F} = -p\log_2(p)-(1-p)\log_2(1-p)
\eeq
where $p=\frac{1}{2}(1+\sqrt{1-|C|^2)}$. For $|C|\approx 1$ we
find
\beq
    E_{\rm F} \approx 1-\frac{1-|C|^2}{\ln2} \ .
    \label{E_F_C}
\eeq

Any resolution of the phase difference requires a minimum variance
$\ip{\Delta \hat N_{\rm Tr}^2}$ in the number of particles
transported from one site to the other. We can relate $E_{\rm F}$
to the variance in particle number using the Heisenberg-Robertson
uncertainty relation for phase and number operators.  In
\ref{HRUR} we find that the optimum strategy gives
\beq
  |C|^2\le
     \frac{4\ip{\Delta \hat N_{\rm Tr}^2}}
     {1+4\ip{\Delta \hat N_{\rm Tr}^2}}
\eeq
Thus, from \erf{E_F_C}, an upper bound for the entanglement of
formation is given approximately by
\beq
  E_{\rm F}\le 1 - \frac{1}
     {4\ip{\Delta \hat N_{\rm Tr}^2}\ln2}
     \label{E_F_optimum}
\eeq
in the limit that $\ip{\Delta \hat N_{\rm Tr}^2}\gg 1$.

As an example, let the ancillary states
$\ket{c(\theta)}_\A\otimes\ket{c(\phi)}_\B$ in \erf{super_m}
approximate two coherent states of not necessarily the same
amplitude, and imagine transporting one of these ancillary modes
from one site to the other to allow an ideal phase-difference
measurement between them. In this case, the variance is given by
$\ip{\Delta \hat N_{\rm Tr}^2}=\ip{\hat N_{\rm Tr}}$ where
$\ip{\hat N_{\rm Tr}}$ is the mean particle number transported
between the sites and so for the optimum strategy $ E_{\rm F} \le
1-1/(4\ip{\hat N_{\rm Tr}}\ln2)$. In fact, a direct calculation of
\erf{C}, using periodic Gaussian distributions to approximate the
phase distributions of the coherent states\cite{PB} and assuming
that the local coherent state has a much larger amplitude than the
one which is transported, gives $|C|^2\approx e^{-1/4\ip{\hat
N_{\rm Tr}}}\approx 1-1/4\ip{\hat N_{\rm Tr}}$ for $\ip{\hat
N_{\rm Tr}}\gg 1$. Thus, from \erf{E_F_C}, using coherent states
to establish a shared reference phase gives the entanglement of
formation as
\beqa
  E_{\rm F} \approx 1-\frac{1}{4\ip{\hat N_{\rm Tr}}\ln2}
\eeqa
for $\ip{\hat N_{\rm Tr}}\gg 1$. This value represents the upper
bound in \erf{E_F_optimum}. Clearly $E_{\rm F}$ approaches 1 ebit
as $\ip{\hat N_{\rm Tr}}$, the mean number transported, increases.

\section{Discussion} \label{discussion}

Only manipulations by local operations and classical communication
are permissible when quantifying the accessible entanglement in a
system. Operations that change local particle number are therefore
forbidden and this gives rise to a local particle-number
superselection rule. This concept underlies the definition of
$E_{\rm P}$, the entanglement of identical
particles.\cite{WiseVacc}  $E_{\rm P}$ quantifies the amount of
accessible entanglement in a system of identical particles, where
the accessibility implies that the entanglement is able to be
transferred to regular quantum registers such as qubits, and be
used as a generic resource in quantum information processing.

In this paper we have shown that the process of transferring the
entanglement of shared particles to quantum register {\em in the
absence of any shared nonlocal resources} necessarily involves
{\em random phase differences between the two sites}. The unknown
nature of these phase differences leads to a reduction in the
transferred entanglement. Any non zero entanglement remaining
after the transfer is constrained to {\em reference-phase
invariant subspaces}. Moreover, the super-additivity of $E_{\rm
P}$ can be attributed to this {\em constraint being weaker for
combined systems} compared to the individual systems. We also
showed that the entanglement can be recovered by establishing a
shared reference phase for the two sites. This operation, however,
requires the transport of particles between the sites, that is, it
is a non local operation. In other words, establishing a shared
reference phase violates the restriction to local operations and
the local superselection rule, and in doing so increases the
accessible entanglement.  We gave a general expression that
relates the transferred entanglement to the variance in the number
of particles transported for the case of a single shared particle.

\section*{Acknowledgements}
We thank Rob Spekkens, Stephen Bartlett and Stephen Barnett for
helpful discussions.

\appendix
\section{} \label{gen_protocol}

We describe here the details of the protocol that transfers the
particle entanglement of the arbitrary $N$-particle state given by
\erf{generalN} into the state \erf{mixedN} of regular quantum
registers. The ket $\ket{\psi_n}_{\rm AB}$ in \erf{generalN}
represents the state of $n$ particles at Alice's site and $N-n$ at
Bob's site, which we write here as
\beqa
  &&\ket{\psi_n}_{\rm AB}=\sum
    d_{u_1^{(n)},u_2^{(n)},\cdots,v_1^{(n)},v_2^{(n)},\cdots}\nn\\
   &&\ \ \times \ket{u_1^{(n)},u_2^{(n)},\cdots}_\A
    \otimes\ket{v_1^{(n)},v_2^{(n)},\cdots}_\B\ .
\eeqa
Here $d_{\cdots}$ are complex amplitudes, $\ket{k_1,k_2,\cdots}_Z$
represents a set of field modes at site $Z\in\{\A,\B\}$ with
corresponding occupations $k_1$, $k_2$, $\cdots$, and the sets of
non-negative integers $u_1^{(n)}$, $u_2^{(n)}$, $\cdots$ and
$v_1^{(n)}$, $v_2^{(n)}$, $\cdots$ have the property that
\beqa
   \sum_{m=0}^N u_m^{(n)} = n,\ \sum_{m=0}^N v_m^{(n)} = N-n\ .
\eeqa
We imagine a corresponding set of regular quantum registers
located at each site and initially in the state
$\aket{0,0,\cdots}_\A\otimes\aket{0,0,\cdots}_\B$.  The system at
Alice's site can be written in part as
\beq
   \cdots\ket{\psi(\theta)}_\A\otimes\ket{u_1^{(n)},u_2^{(n)},\cdots}_\A
        \otimes\aket{0,0,\cdots}_\A\ .
\eeq
Alice performs a unitary operation which transforms her quantum
registers to
\beq
   \cdots\ket{\psi(\theta)}_\A\otimes\ket{u_1^{(n)},u_2^{(n)},\cdots}_\A
      \otimes\aket{u_1^{(n)},u_2^{(n)},\cdots}_\A\ .
\eeq
She then ``hides'' the $n$ shared particles in her truncated phase
state as before; this leaves her system in a state closely
approximated by
\beq
  \cdots e^{-in\theta}\ket{\psi(\theta)}_\A\otimes\ket{0,0,\cdots}_\A
      \otimes\aket{u_1^{(n)},u_2^{(n)},\cdots}_\A\ .
\eeq
Bob repeats these operations at his site.  The end result of
Alice's and Bob's actions is a state of the form
\beqa
  &&\cdots e^{-in\theta}\ket{\psi(\theta)}_\A\otimes\ket{0,0,\cdots}_\A
    \otimes\aket{u_1^{(n)},u_2^{(n)},\cdots}_\A\nonumber\\
  &&\otimes e^{-i(N-n)\phi}\ket{\psi(\theta)}_\B
    \otimes\ket{0,0,\cdots}_\B\otimes
    \aket{v_1^{(n)},v_2^{(n)},\cdots}_\B\ .\ \ \ \ \ \
\eeqa
Including the integrals over the phase angles $\theta$ and $\phi$
and the remaining particle modes, and then tracing over the
particle modes yields the final state of the regular quantum
registers; this is given by \erf{mixedN} with
\beqa
  \aket{\psi_n}_{\rm AB}&=&\sum
    d_{u_1^{(n)},u_2^{(n)},\cdots,v_1^{(n)},v_2^{(n)},\cdots}\nn\\
    \ \ &&\times
    \aket{u_1^{(n)},u_2^{(n)},\cdots}_\A
    \otimes\aket{v_1^{(n)},v_2^{(n)},\cdots}_\B\ .\
\eeqa

\section{}\label{phase_meas}

In this appendix we derive the state of the regular qubits in
\erf{trans1} following an ideal phase-difference measurement of
the ancillary states $\ket{c(\theta)}_\A\otimes\ket{c(\phi)}_\B$
in \erf{super_m}. An ideal phase-difference measurement is
described by the POVM\cite{PB,LV}
\beq
   \hat \Pi^{(-)}(\varphi)  =  \int_{2\pi} \hat
   \Pi_\A(\theta')\otimes\hat
      \Pi_\B(\theta'+\varphi) d\theta'\ ,
\eeq
where $\varphi$ represents the measured value of the phase
difference and $\hat \Pi_Z(\theta)$ is the POVM representing an
ideal measurement of the phase of a field mode at site $Z\in\{\A,
\B\}$:
\beq
  \hat \Pi_Z(\theta)=\frac{1}{2\pi}\sum_{n,m=0}^\infty
  e^{i(n-m)\theta}(\ket{n}\bra{m})_Z\ .
\eeq
The completeness of these POVMs is given by
\beqa
   \int_{2\pi}\hat
      \Pi^{(-)}(\varphi)d\varphi  &=&  \hat\id_\A\otimes\hat\id_\B\ ,\\
  \int_{2\pi}\hat \Pi_Z(\varphi)d\varphi  &=&  \hat\id_Z
\eeqa
where $\hat\id_Z$ is the identity operator for the mode at site
$Z\in\{\A, \B\}$. While it is impossible to realize these
measurements exactly,\footnote{The same can be said, for example,
for position measurements: measurements of position can be made
with arbitrary precision, however, the position POVM
$\ket{x}\bra{x}$, where $\ket{x}$ is an eigenstate of the position
operator, can never be implemented exactly.} nevertheless they can
be implemented, in principle, with arbitrary
precision.\cite{PB,VPB} Consider the full state represented by the
first line of \erf{trans1}:
\beqa
   &&\frac{M+1}{(2\pi)^2}\int_{2\pi}\int_{2\pi}
      \ket{c(\theta)}_{\rm A}\otimes\ket{c(\phi)}_{\rm B}
      \otimes\ket{\psi(\theta)}_\A\otimes
   \ket{0,0}_{\rm AB}\nonumber\\
   &&\ \ \ \otimes\frac{1}{\sqrt{2}}\left(e^{-i\theta}\aket{1,0}_{\rm AB}
   +e^{-i\phi}\aket{0,1}_{\rm AB}\right)\otimes\ket{\psi(\phi)}_\B
   d\theta d\phi\ .\nn\\
\eeqa
Tracing over the shared particle modes and the modes in the
truncated phase states, and using an argument similar to that used
to derive results \erf{I1} and \erf{I2} shows that the state of
the remaining parts of the system is given by
\begin{widetext}
\beq
   \int_{2\pi}\int_{2\pi}
      \left[\ket{c(\theta)}_{\rm A}\otimes\ket{c(\phi)}_{\rm B}
            \otimes\frac{\left(e^{-i\theta}\aket{1,0}_{\rm AB}+
            e^{-i\phi}\aket{0,1}_{\rm AB}\right)}{\sqrt{2}}\right]
      \left[{}_{\rm A}\bra{c(\theta)}\otimes{}_{\rm B}\bra{c(\phi)}
            \otimes\frac{\left({}_{\rm AB}\abra{1,0}e^{i\theta}+
        {}_{\rm AB}\abra{0,1}e^{i\phi}\right)}{\sqrt{2}}\right]
           \frac{d\theta}{2\pi}\frac{d\phi}{2\pi}\ .
        \label{before_meas}
\eeq
\end{widetext}
The state of the regular qubits after an ideal phase-difference
measurement has given the result $\varphi$ is found by forming the
product of $\hat \Pi^{(-)}(\varphi)$ with \erf{before_meas} and
taking the partial trace of the result over the field modes;  we
find this gives the mixed state
\beq
  \frac{1}{2}\Big(\aket{1,0}\abra{1,0}
  +C\aket{1,0}\abra{0,1}
             +C^*\aket{0,1}\abra{1,0}
   +\aket{0,1}\abra{0,1}\Big)_{\rm AB}
\eeq
with probability $1/2\pi$ where
\beq
  C=\int_{2\pi}\int_{2\pi}P_\varphi(\theta,\phi)e^{-i(\theta-\phi)}d\theta
  d\phi\ .
\eeq
Here $P_\varphi(\theta,\phi)$ represents the resolution of the
phase-difference measurement,
\beq
  P_\varphi(\theta,\phi)=\int_{2\pi}
  P_\A(\theta-\theta')P_\B(\phi-\varphi-\theta')\frac{d\theta'}{2\pi}
  \label{P_varphi}
\eeq
where $P_Z(\theta)$ is the canonical phase distribution of the
state $\ket{c(0)}_Z$ for a mode at site $Z\in\{\A, \B\}$, i.e.
$P_Z(\theta)=|\sum_n c_n e^{-in\theta}|^2/2\pi$.

\section{} \label{HRUR}

We derive here the optimum conditions for maximizing the
entanglement of formation \erf{E_F_C} using the
Heisenberg-Robertson uncertainty relations for particle number and
phase operators. We note that the commutator of the
number-difference operator with the cosine of the phase difference
operator can be written as
\beqa
  &&[\hat N_\A-\hat N_\B,
  \cos(\hat\phi_\A-\hat\phi_\B)]\nn\\
  &&=  [\hat N_\A,\cos(\hat\phi_\A-\hat\phi_\B)]
     -[\hat N_\B,\cos(\hat\phi_\A-\hat\phi_\B)]
\eeqa
where $\hat\phi_Z$ and $\hat N_Z$ are the Pegg-Barnett phase
operator\cite{PB} and particle number operator, respectively, for
site $Z\in\{\A, \B\}$, and
\beq
    \cos(\hat\phi_\A-\hat\phi_\B)=
    \frac{e^{i(\hat\phi_\A-\hat\phi_\B)}
    +e^{-i(\hat\phi_\A-\hat\phi_\B)}}{2}\ .
\eeq
It is not difficult to show, using the results and methods in Ref.
\refcite{MUS} (see, in particular p. 32), that
\beqa
   \bra{\Phi}[\hat N_\A,\cos(\hat\phi_\A-\hat\phi_\B)]\ket{\Phi}
   &=&   -\bra{\Phi}[\hat N_\B,\cos(\hat\phi_\A-\hat\phi_\B)]\ket{\Phi}\nonumber\\
   &=&   -i\bra{\Phi}\sin(\hat\phi_\A-\hat\phi_\B)\ket{\Phi}
   \label{N_A}
\eeqa
where
\beq
    \sin(\hat\phi_\A-\hat\phi_\B)
    =\frac{e^{i(\hat\phi_\A-\hat\phi_\B)}-e^{-i(\hat\phi_\A-\hat\phi_\B)}}{2i}
\eeq
and so
\beq
  \bra{\Phi}[\hat N_\A-\hat N_\B,
  \cos(\hat\phi_\A-\hat\phi_\B)]\ket{\Phi}
  =-2i\bra{\Phi}\sin(\hat\phi_\A-\hat\phi_\B)\ket{\Phi}
\eeq
where $\ket{\Phi}$ is a physical state.\cite{PB,MUS}  Similarly
\beq
  \bra{\Phi}[\hat N_\A-\hat N_\B,
  \sin(\hat\phi_\A-\hat\phi_\B)]\ket{\Phi}
  =2i\bra{\Phi}\cos(\hat\phi_\A-\hat\phi_\B)\ket{\Phi}\ .
\eeq
Hence, from Robertson's uncertainty relation\cite{Robertson} we
find
\beqa
  \ip{\Delta^2(\hat N_\A-\hat
     N_\B)}\ip{\Delta^2\cos(\hat\phi_\A-\hat\phi_\B)}
     &\ge&  |\ip{\sin(\hat\phi_\A-\hat\phi_\B)}|^2\ ,\nn\\
     \label{dcos}\\
  \ip{\Delta^2(\hat N_\A-\hat
     N_\B)}\ip{\Delta^2\sin(\hat\phi_\A-\hat\phi_\B)}
     &\ge&  |\ip{\cos(\hat\phi_\A-\hat\phi_\B)}|^2\nn\\
     \label{dsin}
\eeqa
for physical states, where $\ip{\Delta^2 \hat Q}=\ip{\hat
Q^2}-\ip{\hat Q}^2$ is the variance in $\hat Q$.  Adding these
inequalities and using
\beqa
   &&\ip{\Delta^2\cos(\hat\phi_\A-\hat\phi_\B)}
       +\ip{\Delta^2\sin(\hat\phi_\A-\hat\phi_\B)}\nonumber\\
   &&=
   1-\ip{\cos(\hat\phi_\A-\hat\phi_\B)}^2-\ip{\sin(\hat\phi_\A-\hat\phi_\B)}^2\nn\\
   &&= 1-|\ip{e^{i(\hat \phi_{\rm A} - \hat \phi_{\rm B} )}}|^2
\eeqa
gives
\beq
   \ip{\Delta^2(\hat N_\A-\hat N_\B)}\left(1-|C|^2\right)
     \ge |C|^2
     \label{inequal}
\eeq
where, from \erf{C},
\beq
   |C|^2=|\ip{e^{i(\hat \phi_{\rm A} - \hat \phi_{\rm B} )}}|^2\ .
\eeq
Rearranging \erf{inequal} and using $\ip{\Delta^2(\hat N_{\rm
A}-\hat N_{\rm B})} =\ip{\Delta^2 \hat N_{\rm A}}+\ip{\Delta^2
\hat N_{\rm B}}$ for uncorrelated fields gives
\beq
  |C|^2\le
     \frac{\ip{\Delta^2 \hat N_{\rm A}}+\ip{\Delta^2 \hat N_{\rm B}}}
     {1+\ip{\Delta^2 \hat N_{\rm A}}+\ip{\Delta^2 \hat N_{\rm B}}}\ .
     \label{C1}
\eeq
In a similar way, we derive the Heisenberg-Robertson uncertainty
relations:
\beqa
  \ip{\Delta^2 \hat N_Z}\ip{\Delta^2\cos(\hat\phi_\A-\hat\phi_\B)}
     &\ge&  \frac{1}{4}|\ip{\sin(\hat\phi_\A-\hat\phi_\B)}|^2\
     ,\nn\\
     \label{dcos2}\\
  \ip{\Delta^2 \hat N_Z}\ip{\Delta^2\sin(\hat\phi_\A-\hat\phi_\B)}
     &\ge&  \frac{1}{4}|\ip{\cos(\hat\phi_\A-\hat\phi_\B)}|^2\
     ,\nn\\
     \label{dsin2}
\eeqa
using the separate commutators of $\hat N_\A$ and $\hat N_\B$ with
$\cos(\hat\phi_\A-\hat\phi_\B)$ and
$\sin(\hat\phi_\A-\hat\phi_\B)$, and then using \erf{dcos2} and
\erf{dsin2} in place of \erf{dcos} and \erf{dsin} we find that
\beq
  |C|^2\le
     \frac{4\ip{\Delta^2 \hat N_Z}}
     {1+4\ip{\Delta^2 \hat N_Z}}
     \label{C2}
\eeq
for site $Z\in\{\A, \B\}$.

We want to derive the conditions for the optimum situation where
the variance in the number of particles transported is the minimum
for a given value of $E_{\rm F}$. Consider the case where Alice
transports particles to Bob. The entanglement of formation $E_{\rm
F}$ given by \erf{E_F_C} increases monotonically with $|C|^2$.
Comparing \erf{C1} and \erf{C2}, we see that the optimum situation
occurs when $\ip{\Delta^2 \hat N_\B} \ge 3\ip{\Delta^2 \hat N_\A}$
for which the bound on $|C|^2$ is given by \erf{C2} with $Z$ being
the transported mode (i.e. A in this case).

\end{document}